\documentclass[a4paper,11pt]{article}
\usepackage{pos}
\usepackage{braket}
\usepackage{comment}
\usepackage{bbm}

\title{The emergence of expanding space-time in the Lorentzian type IIB matrix model with a novel regularization}
\ShortTitle{The emergence of expanding space-time in the Lorentzian type IIB matrix model}

\author*[a]{Mitsuaki Hirasawa}
\author[b]{Konstantinos N. Anagnostopoulos}
\author[c]{Takehiro Azuma}
\author[d]{Kohta Hatakeyama}
\author[d, e]{Jun Nishimura}
\author[b]{Stratos Papadoudis}
\author[f]{Asato Tsuchiya}

\affiliation[a]{Sezione di Milano Bicocca, Istituto Nazionale di Fisica Nucleare,\\ Piazza della Scienza, 3, I-20126 Milano, Italy}

\affiliation[b]{Physics Department, School of Applied Mathematical and Physical Sciences, National Technical University,\\ Zografou Campus, GR-15780 Athens, Greece}
\affiliation[c]{Institute for Fundamental Sciences, Setsunan University,\\ 17-8 Ikeda Nakamachi, Neyagawa, Osaka, 572-8508, Japan}
\affiliation[d]{KEK Theory Center, High Energy Accelerator Research Organization,\\ 1-1 Oho, Tsukuba, Ibaraki 305-0801, Japan}
\affiliation[e]{Graduate University for Advanced Studies (SOKENDAI),\\ 1-1 Oho, Tsukuba, Ibaraki 305-0801, Japan}
\affiliation[f]{Department of Physics, Shizuoka University,\\ 836 Ohya, Suruga-ku, Shizuoka 422-8529, Japan}

\emailAdd{Mitsuaki.Hirasawa@mib.infn.it}
\emailAdd{konstant@mail.ntua.gr}
\emailAdd{azuma@mpg.setsunan.ac.jp}
\emailAdd{khat@post.kek.jp}
\emailAdd{jnishi@post.kek.jp}
\emailAdd{sp10018@central.ntua.gr}
\emailAdd{tsuchiya.asato@shizuoka.ac.jp}

\abstract{The Lorentzian type IIB matrix model is a promising candidate for a non-perturbative formulation of superstring
theory. However, it was recently found that a Euclidean space-time appears in the conventional large-$N$ limit. In this
work, we study the model with a Lorentz invariant mass term which can be considered as an IR regulator. By performing complex Langevin simulations to overcome the sign problem, we observe the emergence of expanding space-time with Lorentzian signature.}

\FullConference{%
  Corfu Summer Institute 2022 "School and Workshops on Elementary Particle Physics and Gravity",\\
  28 August - 1 October, 2022\\
  Corfu, Greece}

\begin{document}
\begin{flushright}
KEK-TH-2536
\end{flushright}
\maketitle

\section{Introduction}
Superstring theory is considered to be a promising candidate for quantum gravity. One of the remarkable features of this theory is that the dimensionality of the space-time in which the theory is defined is not arbitrary but is determined by the consistency of the theory. Specifically, the theory is consistently defined only in 10D space-time. Therefore, it is important to clarify the relationship between the 10D space-time and our (3+1)D space-time.

One mechanism to explain (3+1)D space-time, in the theory, is the compactification, in which the physics in (3+1)D space-time is determined by the structure of the compactified extra dimension. At the perturbative level, one needs to fix the dimensionality of the extra dimensions to 6 by hand, and one also has huge ambiguity in the structure of the extra dimension. However, it is difficult to construct the extra dimensions explicitly by requiring that the physics in (3+1)D space-time is consistent with the Standard Model at the low-energy scale. Even if this is feasible, it is not clear whether one can choose one of the many possibilities at the perturbative level. Therefore, it is important to study the non-perturbative aspects of superstring theory.

The type IIB matrix model, also called IKKT matrix model, was proposed in Ref. \cite{Ishibashi:1996xs}, and it is one of the promising candidates for a non-perturbative formulation of superstring theory. The model is defined by the partition function
\begin{equation}
    \begin{split}
        Z&=\int dA d\Psi d\bar{\Psi}\ e^{i(S_{\rm b} + S_{\rm f})},\\
        S_{\rm b}&=-\frac{N}{4} {\rm Tr}\left\{ -2[A_0, A_i]^2 + [A_i,A_j]^2 \right\},\\
        S_{\rm f}&=-\frac{N}{2} {\rm Tr}\left\{ \bar{\Psi}_\alpha (\Gamma^\mu)_{\alpha\beta}[A_\mu,\Psi_{\beta}] \right\},
    \end{split}
\end{equation}
where $A_\mu$ $(\mu = 0,1,...,9)$ and $\Psi_\alpha$ $(\alpha=1,2,...,16)$ are $N\times N$ Hermitian matrices, and $\Gamma^\mu$ are the 10D Gamma matrices after the Weyl projection.
This model has $\mathcal{N}=2$ supersymmetry (SUSY), which is the maximal SUSY in 10D space-time.
As a consequence, the model includes the gravitational interaction.
From the action, one can see that there are no space-time coordinates a priori, and that they emerge from the degrees of freedom of this model.
According to the SUSY algebra, a homogeneous shift of the diagonal elements of $A_\mu$ corresponds to the translation in the $\mu$ direction in this model.
As a result, we can interpret the eigenvalues of $A_\mu$ as the space-time coordinates. 

The Euclidean version of this model has SO(10) rotational symmetry, which is spontaneously broken to SO(3). This was shown for the first time using the Gaussian expansion method (GEM) in Refs. \cite{Nishimura:2001sx,Kawai:2002jk,Aoyama:2006rk,Nishimura:2011xy}, and non-perturbative Monte Carlo simulations in Refs. \cite{Anagnostopoulos:2017gos,Anagnostopoulos:2020xai} produced consistent results. The relation between the SO(3) symmetric Euclidean space and our (3+1)D space-time is, however, not clear.
Therefore, it is crucial to investigate the Lorentzian version of the model.

Since the action in the Lorentzian model is complex, the usual Monte Carlo methods are not applicable.
This is the sign problem\footnote{
The Euclidean model also has the same problem, but in that case it is due to the fermionic action.
When the phase of the Pfaffian that arises from the integration of the fermionic degrees of freedom is quenched, there is no SSB \cite{Ambjorn:2000dx,Anagnostopoulos:2013xga,Anagnostopoulos:2015gua}. In order to consider the effect of the dynamics of the fermionic degrees of freedom, the authors in  \cite{Anagnostopoulos:2017gos,Anagnostopoulos:2020xai} used the complex Langevin method. We employ the same method in this work.}, and
it is necessary to deal with it properly to obtain correct results.
The first-principle calculations of the Lorentzian model were done in Refs. \cite{Kim:2011cr, Ito:2013qga, Ito:2013ywa, Ito:2015mxa, Ito:2015mem}, where an approximation was used to avoid the sign problem. Then, the expanding (3+1)D space-time was observed.
However, it was found in Ref. \cite{Aoki:2019tby} that the structure of the expanding space is essentially caused by two points, which implies that the space is not continuous.
The emergence of this singular structure is due to the approximation used to avoid the sign problem.
It was found that this approximation amounts to replacing the  Boltzmann weight $e^{iS}$  by $e^{-\beta S}$, where $\beta$ is a positive constant.
See Ref. \cite{Brahma:2021tkh, Brahma:2022dsd, Steinacker:2023myp, Laliberte:2023bai, Brandenberger:2023ver} for other recent studies on the type IIB matrix model, in which possible applications to cosmology are discussed.

Recently, we have been studying the Lorentzian model without the approximation by using the Complex Langevin Method (CLM) to overcome the sign problem \cite{Hirasawa:2021xeh,Hatakeyama:2021ake,Hatakeyama:2022ybs,Anagnostopoulos:2022dak,Hirasawa:2022qzg}.
In this talk, we report on the current status of our work.

The rest of this paper is organized as follows.
In Sec. 2, we discuss the relationship between the Lorentzian and the Euclidean models.
In Sec. 3, we introduce  a regulator in the Lorentzian model to make it well-defined.
This regulator was also used in the classical analysis in \cite{Hatakeyama:2019jyw}.
In Sec. 4, we explain the CLM and its application to the type IIB matrix model.
The results obtained by the complex Langevin simulations are presented in Sec. 5.
Sec. 6 is devoted to a summary and discussions.

\section{Relationship between the Lorentzian and Euclidean models}
In this section, we explain the relationship between the Lorentzian and the Euclidean models.
For simplicity, here we consider the bosonic model, in which the fermionic contribution is omitted.
The partition function of the model is given by
\begin{equation}
    \begin{split}
        Z&=\int dA\ e^{iS_{\rm b}},\\
        S_{\rm b}&=-\frac{N}{4} {\rm Tr}\left\{ -2[A_0, A_i]^2 + [A_i,A_j]^2 \right\}.
    \end{split}
\end{equation}
Here, we consider a Wick rotation as
\begin{equation}
    \tilde{S}_{\rm b} =-\frac{N}{4} e^{i\frac{\pi}{2}u} {\rm Tr}\left\{ -2e^{-i\pi u}[\tilde{A}_0, \tilde{A}_i]^2 + [\tilde{A}_i,\tilde{A}_j]^2 \right\}\, .
\end{equation}
We rotate both on the world sheet and in the target space at the same time using  one parameter $u$.
Here, $u=0$ corresponds to the Lorentzian model, while $u=1$ corresponds to the Euclidean model.
This Wick rotation is equivalent to the contour deformation
\begin{equation}
    \begin{split}
        \tilde{A}_0 &= e^{i\frac{\pi}{2}u} e^{-i\frac{\pi}{8}u}A_0 = e^{i\frac{3\pi}{8}u}A_0\ ,\\
        \tilde{A}_i &= e^{-i\frac{\pi}{8}u}A_i\ .
    \end{split}
\end{equation}
Note that $e^{-i\frac{\pi}{8}u}$ and $e^{i\frac{\pi}{2}u}$ are the phases of the Wick rotations on the world sheet and in the target space, respectively.

\begin{figure}
\centering
    \includegraphics[width=0.40\hsize]{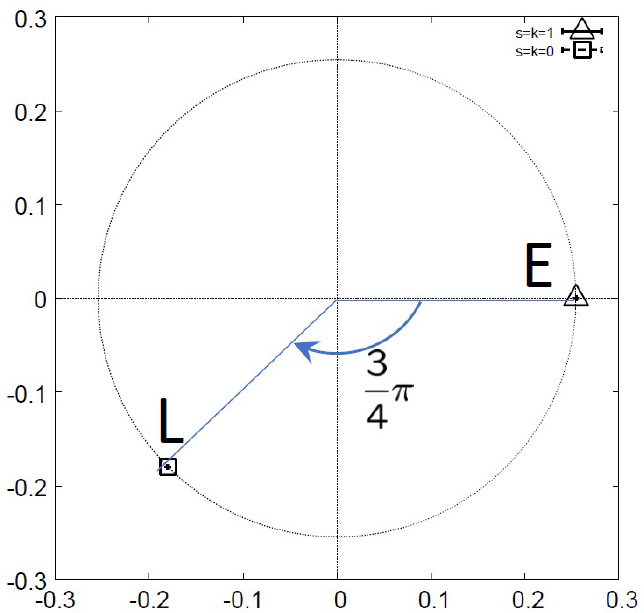}\hspace{3em}
    \includegraphics[width=0.40\hsize]{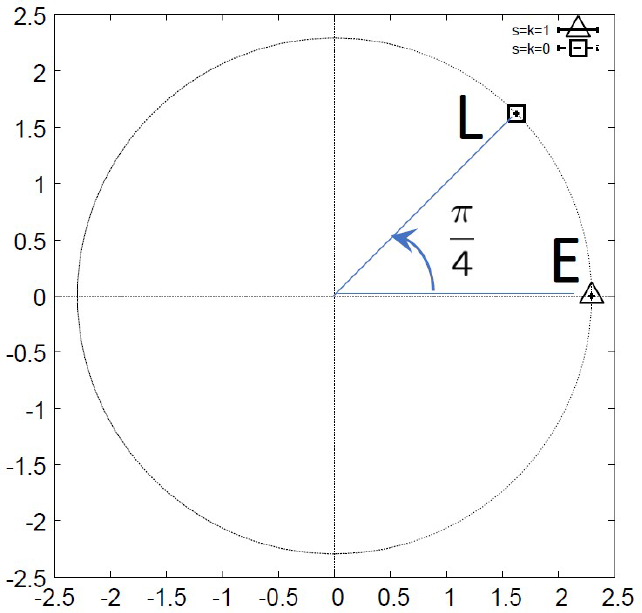}
\caption{The equivalence between the Euclidean and Lorentzian models.
The Left and Right panels represent ${\rm Tr}(A_0)^2$ and ${\rm Tr}(A_i)^2$ in the complex plane.
The data points with ``L" and ``E" represent the expectation values obtained by simulations of the Lorentzian and Euclidean models, respectively.}
\label{fig:equiv_EL}
\end{figure}

Cauchy's theorem says that the expectation value of any observable $\braket{\mathcal{O}(e^{-i\frac{3\pi}{8}u}\tilde{A}_0, e^{i\frac{\pi}{8}u}\tilde{A}_i)}_u$ is independent of $u$ under the contour deformation.
Therefore, the following relations hold:
\begin{equation}
\begin{split}
    \Braket{\frac{1}{N} {\rm Tr} (A_0)^2}_{\rm L} &= e^{-i\frac{3\pi}{4}} \Braket{\frac{1}{N} {\rm Tr} (\tilde{A}_0)^2}_{\rm E},\\
    \Braket{\frac{1}{N} {\rm Tr} (A_i)^2}_{\rm L} &= e^{i\frac{\pi}{4}} \Braket{\frac{1}{N} {\rm Tr} (\tilde{A}_i)^2}_{\rm E},
\end{split}
\label{eq:Cauchys_theore}
\end{equation}
where $\Braket{\cdot}_{\rm L}$ and $\Braket{\cdot}_{\rm E}$ are the expectation values in the Lorentzian and Euclidean models, respectively.
In other words, the Lorentzian and Euclidean models are equivalent to each other under the contour deformation.
We have confirmed this relation by simulations (see. Fig~\ref{fig:equiv_EL}).

\section{Regularization of the Lorentzian model}
Since the partition function of the Lorentzian model is not absolutely convergent as it is, we need to introduce a regularization.
Here, we use the following mass term as an IR regulator
\begin{equation}
    S_{\rm \gamma} = \frac{1}{2}N\gamma \left\{ {\rm Tr} (A_0)^2 - {\rm Tr} (A_i)^2 \right\},
\end{equation}
where $\gamma$ is a mass parameter.
This mass term is invariant under a Lorentz transformation in the target space-time.

The model with this mass term has been studied at the classical and perturbative levels in Refs. \cite{Steinacker:2017vqw, Steinacker:2017bhb, Sperling:2018xrm, Sperling:2019xar, Steinacker:2019dii, Steinacker:2019awe, Steinacker:2019fcb, Hatakeyama:2019jyw, Steinacker:2020xph, Fredenhagen:2021bnw, Steinacker:2021yxt, Asano:2021phy, Karczmarek:2022ejn, Battista:2022hqn}.
In Ref. \cite{Hatakeyama:2019jyw}, it was found that the typical classical solutions of the model with this mass term have an expanding space at $\gamma>0$, although the dimensionality is not determined by the classical analysis.
This result motivates us to perform a first-principle calculation of the model with this mass term.

\section{Complex Langevin simulations}
In order to study the time evolution in Sec. \ref{Sec:results}, we choose an SU($N$) basis, where the temporal matrix $A_0$ is diagonalized as
\begin{equation}
    \label{kna01}
    A_0 = {\rm diag}\left(\alpha_1, \alpha_2, ..., \alpha_N \right),\hspace{2em} \alpha_1 \le \alpha_2 \le ... \le \alpha_N.
\end{equation}
The following change of variables, first introduced in Ref. \cite{Nishimura:2019qal}, make the ordering in Eq. (\ref{kna01}) explicit:
\begin{equation}
    \alpha_1 = 0,\hspace{1em} \alpha_2 = e^{\tau_1},\hspace{1em} \alpha_3 = e^{\tau_1} + e^{\tau_2} ,\hspace{1em} ...,\hspace{1em} \alpha_N = \sum_{a=1}^{N-1}e^{\tau_a}\, .
\end{equation}
The choice $\alpha_1 = 0$ is made by using the shift symmetry $A_0 \to A_0 + c \mathbbm{1}$.

In this work, we use the complex Langevin method (CLM) \cite{Parisi:1983mgm,Klauder:1983sp} to overcome the sign problem.
In this method, the number of degrees of freedom is doubled by ``complexifying" the dynamical variables as
\begin{equation}
\begin{split}
    \tau_a \in \mathbb{R}  &\to  \tau_a \in \mathbb{C},\\
    A_i: {\rm Hermitian\ matrices}  &\to A_i: {\rm general\ complex\ matrices}.
\end{split}
\end{equation}
We generate configurations by using the complex Langevin equations
\begin{equation}
\begin{split}
    \frac{d\tau_a}{dt_{\rm L}} &= - \frac{\partial S}{\partial \tau_a} + \eta_a(t_{\rm L}),\\
    \frac{d(A_i)_{ab}}{dt_{\rm L}} &= - \frac{\partial S}{\partial (A_i)_{ba}} + (\eta_i)_{ab}(t_{\rm L}),
\end{split}
\end{equation}
where $t_{\rm L}$ is the so-called Langevin time, and $\eta_a(t_{\rm L})$ and $(\eta_i)_{ab}(t_{\rm L})$ are the Gaussian noise with the probability distribution
\begin{equation}
\begin{split}
    P\left(\eta_a(t_{\rm L})\right) &\propto \exp \left( -\frac{1}{4} \int dt\ \sum_a (\eta_a(t_{\rm L}))^2 \right), \\
    P\left((\eta_i)_{ab}(t_{\rm L})\right) &\propto \exp \left( -\frac{1}{4} \int dt\ {\rm Tr} \left(\eta_i(t_{\rm L})\right)^2 \right).
\end{split}
\end{equation}
Note that the Langevin equation must be extended to the complexified dynamical variables in a holomorphic way.

It is known that the CLM sometimes converges to wrong solutions. This is called the wrong convergence problem.
Fortunately, a practical criterion for the correct convergence was found recently in Ref. \cite{Nagata:2016vkn}.
The criterion says that the results are correct when the probability distribution of the drift term decays exponentially or faster.

When we consider the fermionic contribution, the inverse of the Dirac operator appears in the drift force.
If the Dirac operator has near zero eigenvalues, we have  the singular drift problem, and the CLM suffers from  the wrong convergence problem.
We avoid this problem by adding a SUSY-breaking fermionic mass term
\begin{equation}
    S_{m_{\rm f}} = iNm_{\rm f} {\rm Tr} \left[ \bar{\Psi}_\alpha (\Gamma_7\Gamma_8^\dagger\Gamma_9)_{\alpha\beta}\Psi_\beta \right]\, ,
\end{equation}
used in the studies of the Euclidean model \cite{Anagnostopoulos:2017gos,Anagnostopoulos:2020xai}.
The original model is obtained after an  $m_{\rm f} \to 0$ extrapolation.

For  some values of $\gamma$ the CLM can become unstable, and  to stabilize the simulation, we perform the redefinition
\begin{equation}
    A_i \to \frac{A_i+\epsilon A_i^\dagger}{1+\epsilon}
\label{eq:dynamical_stabilization}
\end{equation}
after each Langevin step. This procedure is similar to the dynamical stabilization used in lattice QCD simulations \cite{Attanasio:2018rtq}, 
and its effect is expected to be small when the $A_i$ are near Hermitian.

\section{Results}
\label{Sec:results}

All of the results presented in this paper are obtained through simulations in which the criterion for correct convergence is satisfied.
We fix the matrix size to $N=64$, and  use $\epsilon=0.01$ for the dynamical stabilization (\ref{eq:dynamical_stabilization}).

\begin{figure}
    \centering
    \includegraphics[height=14em]{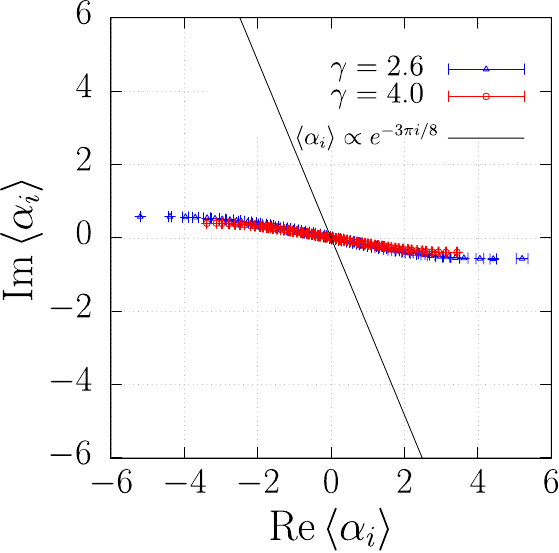}\hspace{2em}
    \includegraphics[height=14em]{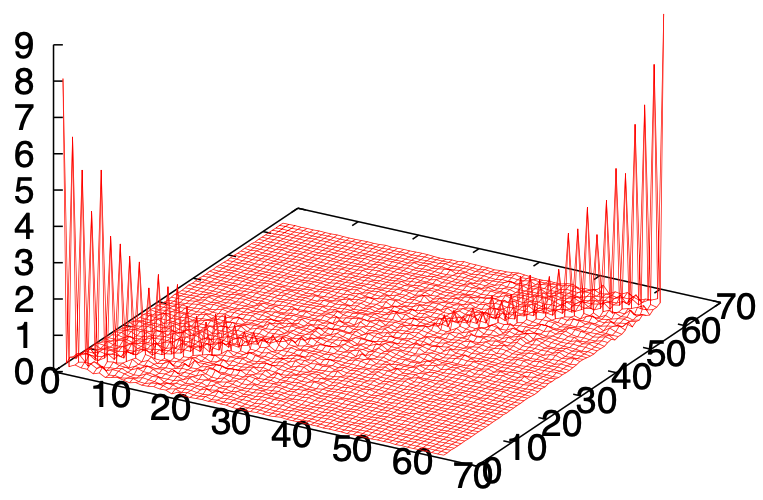}
    \caption{(Left) The $\alpha$-distribution in the complex plane. If the equivalence (\ref{eq:Cauchys_theore}) holds, the $\alpha$'s distribute on the solid line.
    We plot the results obtained by simulations for three different values of $\gamma$, while $N=64$, $m_{\rm f}=10$ and $\epsilon=0.01$ are kept fixed.
    $\alpha_i$ become almost real at $\gamma \ge 2.6$.
    (Right) The 3D plot of $\mathcal{A}_{pq}$ at $\gamma=4$, where the $x$ and $y$ axes represent $p$ and $q$, respectively. $\mathcal{A}_{pq}$ is large when $|p-q|$ is small, while $\mathcal{A}_{pq}$ is small when $|p-q|$ is large.}
    \label{fig:alpha_Apq}
\end{figure}

In order to see whether the emergent time is real, we compute the eigenvalues $\alpha_a$ of $A_0$.
In Fig.~\ref{fig:alpha_Apq} (Left), we plot the eigenvalues $\alpha_a$ in the complex plane.
When $2.6 \le \gamma \le 4.0$, the model is in the real-time phase, where $\alpha_{a+1}-\alpha_a$ is almost real at late times.
At smaller $\gamma$, the $\alpha_a$-distribution becomes wider in the real direction.

We define the matrix
\begin{equation}
    \mathcal{A}_{pq} = \frac{1}{9} \sum_{i=1}^9 \left| (A_i)_{pq} \right|^2.
\end{equation}
In Fig.~\ref{fig:alpha_Apq} (Right), we plot $\mathcal{A}_{pq}$ against $p$ and $q$.
$\mathcal{A}_{pq}$ is large for small $|p-q|$, and drops fast to very small values with increasing $|p-q|$, showing that
the matrices $A_i$ have a band diagonal structure.
We define the band width $n$ such that $\mathcal{A}_{pq} \approx 0$ when $|p-q| > n$.
In this work, we choose $n=12$.

The appearance of the band diagonal structure motivates us to define the time and the block matrices that describe the state of the universe at that time as follows:
Time is defined using the average of $n$ diagonal elements
\begin{equation}
    t_a = \sum_{i=1}^a |\bar{\alpha}_i - \bar{\alpha}_{i-1}|\, , \quad a = 1, 2, \ldots, N-n\, ,
\end{equation}
where $\bar{\alpha}_i$ is an average of the $\alpha$'s in the $i$-th block:
\begin{equation}
    \bar{\alpha}_i = \frac{1}{n} \sum_{\nu=1}^{n} \alpha_{i+\nu}\, , \quad i=0, 1, \ldots, N-n\, .
\end{equation}
We define the $n \times n$ block matrices within the spatial matrices as
\begin{equation}
    \left( \bar{A}_i \right)_{kl}(t_a) = \left( A_i \right)_{(k+a-1)(l+a-1)}\, ,\quad k,l=1,2,\ldots,n\, .
\end{equation}
We interpret these block matrices to represent the state of the universe at $t_a$.
In the following, we omit the index of $t_a$ and use $t$ for simplicity.

\begin{figure}
    \centering
    \includegraphics[width=0.49\hsize]{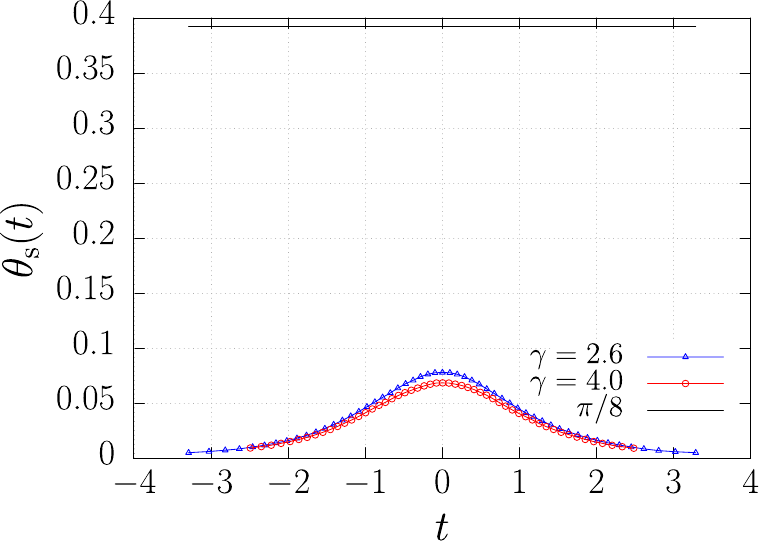}
    \caption{The $\theta_{\rm s}(t)$ is plotted against time at $m_{\rm f}=10$ for $\gamma=2.6$ and 4.
    When the equivalence (\ref{eq:Cauchys_theore}) holds,  the data points lie on the horizontal solid line $\theta_{\rm s}=\pi/8$.
    We see that $\theta_{\rm s}(t)$ approaches 0 at late times.
    }
    \label{fig:theta_s}
\end{figure}

In order to check whether the space is real or complex, we define the phase $\theta_{\rm s}(t)$ as
\begin{equation}
    {\rm tr}\left( \bar{A}_i(t) \right)^2 = e^{2i\theta_{\rm s}(t)}\left| {\rm tr}\left( \bar{A}_i(t) \right)^2 \right|.
\end{equation}
We plot $\theta_{\rm s}(t)$ against $t$ at $\gamma=2.6$ in Fig.~\ref{fig:theta_s}.
At $\gamma=2.6$, the $\alpha$'s are almost real, and the phase $\theta_s(t)$ approaches 0 at late times.
Therefore, at late times, we obtain real space-time.

For the purpose of studying the SSB of the spatial SO(9) symmetry, we define the ``moment of inertia tensor" as
\begin{equation}
    T_{ij}(t) = {\rm tr}\left( X_i(t) X_j(t) \right),
\end{equation}
where $X_i(t)$ are the Hermitian matrices 
\begin{equation}
    X_i(t) = \frac{\bar{A}_i(t) + \bar{A}_i^\dagger(t)}{2}.
\label{eq:hermitization}
\end{equation}
As $\theta_{\rm s}(t)$ is near zero, particularly at late times, the block matrices are near Hermitian.
Therefore, the procedure (\ref{eq:hermitization}) is justified, even though the CLM only allows the calculation of holomorphic observables.

\begin{figure}
    \centering
    \includegraphics[width=0.49\hsize]{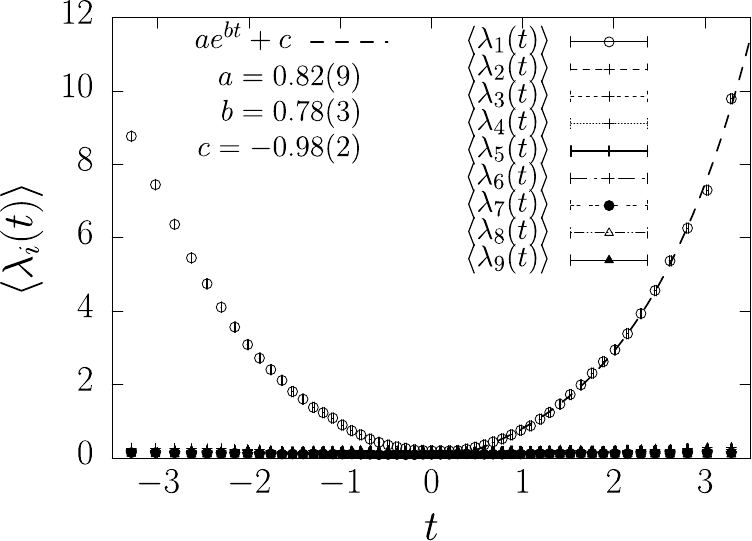}
    \includegraphics[width=0.49\hsize]{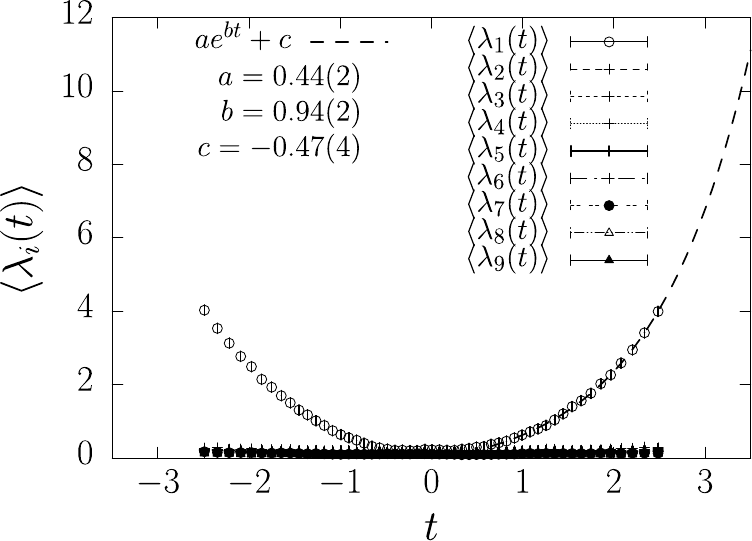}
    \caption{The eigenvalues $\lambda_i(t)$  of $T_{ij}(t)$ are plotted against time at $m_{\rm f}=5$.
    The Left and Right panels show results for $\gamma=2.6$ and $\gamma=4.0$, respectively.
    The dotted lines are obtained by exponential fittings.
    In both cases, one out of nine eigenvalues grows exponentially with time.
    The extent of time becomes larger at smaller $\gamma$.}
    \label{fig:Tij_gam}
\end{figure}

In Fig.~\ref{fig:Tij_gam}, we plot the eigenvalues of $T_{ij}(t)$ as a function of time at $\gamma=2.6$ and $4$.
As we can see, the eigenvalues are almost degenerate around $t=0$.\footnote{
Due to the finite-$N$ effects, the 9 eigenvalues are not exactly degenerate.
}
At some point in time, one out of nine eigenvalues starts to grow exponentially.
Thus, a 1-dimensional space expands exponentially.
The extent of time is larger at smaller $\gamma$.

\begin{figure}
    \centering
    \includegraphics[width=0.49\hsize]{Tij_N64_mf10_gam2.6_2.pdf}
    \includegraphics[width=0.49\hsize]{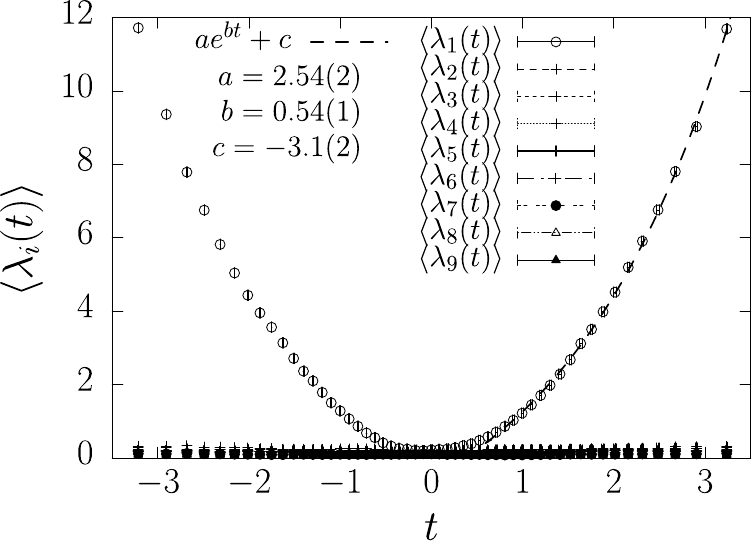}
    \caption{The eigenvalues $\lambda_i(t)$  of $T_{ij}(t)$ are plotted against time at $\gamma=2.6$.
    The Left and Right panels show results for $m_{\rm f}=10$ and $m_{\rm f}=5$, respectively.
    The dotted lines represent fits to the exponential behavior.
    The expansion of space gets more pronounced at smaller $m_{\rm f}$.}
    \label{fig:Tij_mf}
\end{figure}

In order to see the $m_{\rm f}$ dependence, we compare the results at $m_{\rm f}=10$ and $5$ in Fig.~\ref{fig:Tij_mf}.
The expansion of space becomes more pronounced as $m_{\rm f}$ decreases.
The reason for this is that the SUSY effects weaken the attractive force between space-time eigenvalues.

\section{Summary and discussions}
We have conducted an investigation into the emergence of space-time in the type IIB matrix model.
We primarily focused on the Lorentzian version of the model as the Euclidean version of the model revealed the SSB of the SO(10) to SO(3), and the connection between the emerging 3D space and our (3+1)D universe was not clear \cite{Ambjorn:2000dx,Kawai:2002jk,Aoyama:2006rk,Nishimura:2011xy, Anagnostopoulos:2013xga,Anagnostopoulos:2015gua,Anagnostopoulos:2017gos,Anagnostopoulos:2020xai}.

We have used a Lorentz invariant mass term as a regulator, which breaks the equivalence between the Lorentzian and the Euclidean models.
This was motivated by the results in Ref. \cite{Hatakeyama:2019jyw}, where the typical classical solutions with positive mass term ($\gamma > 0$) represent an expanding space, whose dimensionality is not fixed at the classical level.

We employed the CLM to overcome the sign problem, and found that real space-time appears for $2.6 \le \gamma \le 4$.
The SO(9) rotational symmetry of space breaks spontaneously, and one spatial dimension expands exponentially with time.
This expansion becomes stronger as the fermionic mass $m_{\rm f}$ decreases.

The reason why the 1-dimensional expanding space appears may be explained as follows.
If we ignore the fermionic contribution, the configurations that minimize $-{\rm Tr}[A_i,A_j]^2$ are dominant.
Therefore, configurations in which the expanding space has small dimensionality are favored.
Particularly, when only one out of the nine matrices is large and the remaining nine are almost zero,  $-{\rm Tr}[A_i,A_j]^2$ acquires the minimum value ($-{\rm Tr}[A_i,A_j]^2=0$).
In Refs. \cite{Krauth:1998xh, Nishimura:2000ds}, the effect of the Pfaffian of the Dirac matrix was studied, and it was found that the Pfaffian becomes 0 when only 2 out of 10 matrices are nonzero. Then, the appearance of less than 2-dimensional expanding space must be highly suppressed. Therefore, we conclude that  $m_{\rm f}=5$ is not small enough to make this effect dominant, and we expect that the emergence of the expanding 3-dimensional space will occur by further decreasing  $m_{\rm f}$.

\section*{Acknowledgements}
T.\;A., K.\;H. and A.\;T. were supported in part by Grant-in-Aid (Nos.17K05425, 19J10002, and 18K03614, 21K03532, respectively)
from Japan Society for the Promotion of Science.  This research was supported by MEXT as ``Program for Promoting Researches on the
Supercomputer Fugaku'' (Simulation for basic science: from fundamental laws of particles to creation of nuclei, JPMXP1020200105) and JICFuS.
This work used computational resources of supercomputer Fugaku provided by the RIKEN Center for Computational Science (Project ID: hp210165, hp220174), and Oakbridge-CX provided by the University of Tokyo (Project IDs: hp200106, hp200130, hp210094, hp220074) through the HPCI System Research Project.
Numerical computations were also carried out on PC clusters in KEK Computing Research Center.  
This work was also supported by computational time granted by the Greek Research and Technology Network (GRNET) in the National HPC facility ARIS, 
under the project IDs LIIB and LIIB2.
K.\;N.\;A and S.\;P. were supported in part by a Program of Basic Research PEVE 2020 (No. 65228700) of the National Technical University of Athens.

\bibliographystyle{JHEP}
\bibliography{bib.bib}
\end{document}